\documentclass[a4paper,11pt]{article}
\usepackage{amsmath,amssymb,amsfonts}
\usepackage{geometry}
\usepackage{setspace}

\title{\textbf{An Entropy-Based Criterion for Geodesic Incompleteness}}

\author{Rohit Dhormare\thanks{Email: dhormaretheoreticalphysics@proton.me} \\
\textit{Dr. Babasaheb Ambedkar Marathwada University, India}}

\date{}
\begin{document}

\maketitle

\begin{abstract}
We develop an entropy-based formulation of gravitational focusing in Lorentzian geometry by associating a scalar entropy functional to timelike geodesic congruences. Rewriting the Raychaudhuri equation as an entropy production law, we show that entropy accumulation governs the evolution of the expansion scalar and drives finite-time focusing. We prove that when the integrated entropy exceeds a critical threshold set by the initial expansion, geodesic incompleteness necessarily follows. This result provides a quantitative refinement of the classical singularity theorems, replacing qualitative geometric conditions with an explicit entropy criterion. Our approach offers a thermodynamic interpretation of gravitational collapse and suggests that singularities arise as endpoints of irreversible entropy growth in spacetime.
\end{abstract}

\section{Introduction}

The study of singularities occupies a central position in general relativity. The celebrated singularity theorems of Hawking and Penrose~\cite{HawkingPenrose1970,Penrose1965} establish that, under suitable geometric, causal, and energy conditions, spacetime must be geodesically incomplete. These results rely fundamentally on the behavior of geodesic congruences as governed by the Raychaudhuri equation~\cite{Raychaudhuri1955,Wald1984}, which encodes the influence of spacetime curvature on the focusing of nearby geodesics.

For a timelike geodesic congruence with tangent vector field $u^\mu$, the Raychaudhuri equation takes the form:
\begin{equation}
\frac{d\theta}{d\tau} = -\frac{1}{3}\theta^2 - \sigma_{\mu\nu}\sigma^{\mu\nu} - R_{\mu\nu}u^\mu u^\nu,
\label{eq:raychaudhuri}
\end{equation}
where $\theta$ is the expansion scalar, $\sigma_{\mu\nu}$ is the shear tensor, and $R_{\mu\nu}$ is the Ricci tensor. Under the strong energy condition,
\begin{equation}
R_{\mu\nu}u^\mu u^\nu \geq 0,
\label{eq:sec}
\end{equation}
all terms on the right-hand side are non-negative, implying that initially converging congruences ($\theta < 0$) must develop caustics within finite proper time. This focusing mechanism forms the backbone of the Hawking--Penrose singularity theorems.

Despite their profound significance, these theorems are inherently qualitative. They assert the inevitability of geodesic incompleteness but do not provide a quantitative description of the dynamical process leading to collapse. In particular, they do not measure the rate at which focusing occurs, nor do they identify a scalar quantity that accumulates along geodesics and drives the formation of singularities.

In parallel developments, entropy functionals have played a central role in geometric analysis and mathematical physics, particularly in the study of curvature flows and monotonicity formulas. Motivated by these ideas, as well as by the thermodynamic interpretation of gravitational phenomena (see, e.g.,~\cite{Jacobson1995,Padmanabhan2010}), it is natural to ask whether a similar entropy-like quantity can be associated with geodesic congruences in Lorentzian geometry.

In this work, we introduce an entropy density associated with a timelike geodesic congruence:
\begin{equation}
\rho_S = R_{\mu\nu}u^\mu u^\nu + \sigma_{\mu\nu}\sigma^{\mu\nu} + \frac{1}{3}\theta^2,
\label{eq:entropy_density}
\end{equation}
which combines the contributions of Ricci focusing, anisotropic deformation, and isotropic expansion. This quantity is manifestly non-negative under the strong energy condition and naturally encodes the total rate of geometric deformation of infinitesimal volume elements transported along the congruence.

A key observation is that the Raychaudhuri equation can be rewritten in the form:
\begin{equation}
\frac{d\theta}{d\tau} = -\rho_S,
\label{eq:entropy_law}
\end{equation}
which admits a direct interpretation as an entropy production law along the flow. This suggests that gravitational focusing may be viewed as an irreversible process driven by the accumulation of geometric entropy.

Building on this observation, we define a global entropy functional:
\begin{equation}
S(\tau) = \int_{\Sigma_\tau} \rho_S \, d\Sigma,
\label{eq:entropy_functional}
\end{equation}
where $\Sigma_\tau$ is a spacelike hypersurface. We then show that the accumulation of entropy along geodesics provides a quantitative criterion for singularity formation. In particular, we establish that when the integrated entropy exceeds a critical threshold determined by the initial expansion, the congruence must focus within finite proper time, leading to geodesic incompleteness.

This perspective provides a refinement of the classical singularity theorems by replacing purely geometric conditions with an explicit scalar measure of collapse. Moreover, it offers a thermodynamic interpretation of gravitational dynamics in which singularities arise as endpoints of entropy accumulation.

The structure of the paper is as follows. In Section 2, we introduce the geometric setup and define the entropy density and associated functional. In Section 3, we derive the entropy evolution identity and establish monotonicity properties. Section 4 contains the main singularity theorem based on entropy accumulation, together with quantitative bounds. In Section 5, we discuss the physical interpretation and relation to classical results. We conclude with remarks on possible extensions and connections to broader developments in gravitational thermodynamics.
\section{Geometric Setup}

Let $(M,g)$ be a smooth, four-dimensional, globally hyperbolic Lorentzian manifold with signature $(-,+,+,+)$. Global hyperbolicity ensures the existence of a foliation of $M$ by spacelike Cauchy hypersurfaces $\{\Sigma_\tau\}_{\tau \in \mathbb{R}}$, providing a well-posed causal structure and allowing a consistent evolution along timelike directions.

Let $u^\mu$ be a smooth, future-directed timelike vector field generating a congruence of affinely parametrized geodesics:
\begin{equation}
u^\nu \nabla_\nu u^\mu = 0, \qquad u^\mu u_\mu = -1.
\label{eq:geodesic_condition}
\end{equation}
Here $\tau$ denotes proper time along the integral curves of $u^\mu$.

\subsection{Kinematical Decomposition}

The covariant derivative of $u^\mu$ admits the standard irreducible decomposition:
\begin{equation}
\nabla_\nu u_\mu = \frac{1}{3}\theta h_{\mu\nu} + \sigma_{\mu\nu} + \omega_{\mu\nu},
\label{eq:decomposition}
\end{equation}
where:

\begin{itemize}
\item $\theta = \nabla_\mu u^\mu$ is the expansion scalar, measuring the fractional rate of change of an infinitesimal volume element along the flow,

\item $\sigma_{\mu\nu}$ is the shear tensor, defined by
\begin{equation}
\sigma_{\mu\nu} = \nabla_{(\nu} u_{\mu)} - \frac{1}{3}\theta h_{\mu\nu},
\label{eq:shear}
\end{equation}
which is symmetric, trace-free, and orthogonal to $u^\mu$,

\item $\omega_{\mu\nu} = \nabla_{[\nu} u_{\mu]}$ is the vorticity tensor, representing the local rotation of the congruence.
\end{itemize}

The tensor
\begin{equation}
h_{\mu\nu} = g_{\mu\nu} + u_\mu u_\nu
\label{eq:projection_tensor}
\end{equation}
is the projection tensor onto the subspace orthogonal to $u^\mu$ and induces a Riemannian metric on hypersurfaces orthogonal to the flow.

\subsection{Irrotational Congruences}

In this work, we restrict attention to irrotational congruences:
\begin{equation}
\omega_{\mu\nu} = 0.
\label{eq:irrotational}
\end{equation}

By Frobenius’ theorem, this condition implies that the congruence is hypersurface-orthogonal, so that $u^\mu$ is normal to a family of spacelike hypersurfaces $\Sigma_\tau$. This assumption simplifies the dynamics and is natural in the context of gravitational collapse and cosmological models.

\subsection{Raychaudhuri Equation}

The evolution of the expansion scalar $\theta$ along the flow is governed by the Raychaudhuri equation:
\begin{equation}
\frac{d\theta}{d\tau} = u^\mu \nabla_\mu \theta = -\frac{1}{3}\theta^2 - \sigma_{\mu\nu}\sigma^{\mu\nu} - R_{\mu\nu}u^\mu u^\nu,
\label{eq:raychaudhuri_section2}
\end{equation}
where $R_{\mu\nu}$ is the Ricci tensor.

Each term on the right-hand side has a clear geometric interpretation:

\begin{itemize}
\item The term $-\frac{1}{3}\theta^2$ represents nonlinear self-interaction of the expansion and leads to amplification of focusing when $\theta < 0$,

\item The shear term $-\sigma_{\mu\nu}\sigma^{\mu\nu}$ is non-positive and describes anisotropic distortion of the congruence, contributing to focusing,

\item The curvature term $-R_{\mu\nu}u^\mu u^\nu$ encodes the effect of matter through the Einstein equations.
\end{itemize}

Under the strong energy condition:
\begin{equation}
R_{\mu\nu}u^\mu u^\nu \geq 0,
\label{eq:sec_section2}
\end{equation}
all terms on the right-hand side are non-positive, implying:
\begin{equation}
\frac{d\theta}{d\tau} \leq 0.
\label{eq:theta_monotonic}
\end{equation}

Thus, the expansion scalar is monotonically decreasing along the flow. In particular, if $\theta$ is initially negative, it will diverge to $-\infty$ within finite proper time, leading to the formation of caustics and the breakdown of the congruence.

\subsection{Geometric Interpretation}

The Raychaudhuri equation provides a precise description of how spacetime curvature influences the evolution of geodesic congruences. It shows that both matter (via the Ricci tensor) and intrinsic distortions (via shear) act to focus geodesics.

In the framework developed in this work, the right-hand side of the Raychaudhuri equation will be reinterpreted as an effective entropy production term. This perspective allows us to reformulate the focusing mechanism in thermodynamic terms, where singularities arise as a consequence of accumulated geometric entropy along the flow.
\section{Entropy Density and Functional}

\subsection{Entropy Density}

We introduce a scalar quantity associated with a timelike geodesic congruence that captures the total geometric deformation of infinitesimal volume elements transported along the flow.

\textbf{[Entropy Density]} Let $u^\mu$ be a timelike geodesic congruence. The associated entropy density is defined by:
\begin{equation}
\rho_S = R_{\mu\nu}u^\mu u^\nu + \sigma_{\mu\nu}\sigma^{\mu\nu} + \frac{1}{3}\theta^2.
\label{eq:entropy_density_section3}
\end{equation}

This definition combines three fundamental contributions:

\begin{itemize}
\item The Ricci term $R_{\mu\nu}u^\mu u^\nu$, which represents curvature-induced focusing and encodes the influence of matter via the Einstein field equations,

\item The shear invariant $\sigma_{\mu\nu}\sigma^{\mu\nu}$, which measures anisotropic distortion of the congruence,

\item The expansion term $\frac{1}{3}\theta^2$, which quantifies isotropic volume change.
\end{itemize}

\subsection{Positivity and Regularity}

Under the strong energy condition:
\begin{equation}
R_{\mu\nu}u^\mu u^\nu \geq 0,
\label{eq:sec_section3}
\end{equation}
and since both $\sigma_{\mu\nu}\sigma^{\mu\nu}$ and $\theta^2$ are non-negative, we immediately obtain:
\begin{equation}
\rho_S \geq 0.
\label{eq:rho_positive}
\end{equation}

Thus, $\rho_S$ defines a non-negative scalar field on spacetime, which may be interpreted as a local density of geometric entropy.

\textbf{Remark.} The non-negativity of $\rho_S$ is crucial for the monotonicity properties derived in later sections and plays a role analogous to the positivity of entropy production in thermodynamic systems.

\subsection{Control of Kinematical Quantities}

A key feature of $\rho_S$ is that it provides pointwise control over the kinematical invariants of the congruence. In particular, we have:
\begin{equation}
\sigma_{\mu\nu}\sigma^{\mu\nu} \leq \rho_S, \qquad \frac{1}{3}\theta^2 \leq \rho_S.
\label{eq:kinematic_bounds}
\end{equation}

Consequently, on any domain $U \subset M$, one obtains bounds of the form:
\begin{equation}
\|\sigma\|_{L^\infty(U)} \leq C \|\rho_S\|^{1/2}_{L^\infty(U)}, \qquad |\theta| \leq \sqrt{3}\,\rho_S^{1/2},
\label{eq:linfty_bounds}
\end{equation}
for some constant $C > 0$ depending on the choice of frame.

These estimates will play an important role in relating entropy bounds to curvature constraints and focusing behavior.

\subsection{Geometric Interpretation}

The entropy density $\rho_S$ measures the total rate of deformation of infinitesimal volume elements along the congruence. More precisely:

\begin{itemize}
\item The Ricci term governs convergence due to matter fields,

\item The shear term captures anisotropic distortion, which leads to irreversible mixing of neighboring geodesics,

\item The expansion term describes isotropic contraction or expansion.
\end{itemize}

From this perspective, $\rho_S$ encodes both reversible and irreversible aspects of the geometric evolution, and can be viewed as a measure of the deviation from uniform geodesic flow.

\subsection{Entropy Functional}

Let $\{\Sigma_\tau\}$ be a foliation of spacetime by spacelike hypersurfaces orthogonal to $u^\mu$, with induced metric $h_{\mu\nu}$ and volume element:
\begin{equation}
d\Sigma = \sqrt{h}\, d^3x.
\label{eq:volume_element}
\end{equation}

\textbf{[Entropy Functional]} The entropy functional associated with the congruence is defined by:
\begin{equation}
S(\tau) = \int_{\Sigma_\tau} \rho_S \, d\Sigma.
\label{eq:entropy_functional_section3}
\end{equation}

This functional represents the total geometric entropy contained in the hypersurface $\Sigma_\tau$.

\subsection{Regularity Assumptions}

In what follows, we assume sufficient regularity of the geometric fields so that:
\begin{equation}
\rho_S \in L^\infty(U)
\label{eq:regularity}
\end{equation}
for any compact domain $U \subset M$, and that the congruence admits smooth evolution along $\tau$.

These assumptions ensure that the entropy functional is well-defined and allows differentiation under the integral sign when analyzing its evolution.

\subsection{Interpretation as a Thermodynamic Quantity}

The structure of $\rho_S$ suggests a natural analogy with non-equilibrium thermodynamics:

\begin{itemize}
\item $\sigma_{\mu\nu}\sigma^{\mu\nu}$ corresponds to shear viscosity,

\item $\theta^2$ corresponds to bulk viscosity,

\item $R_{\mu\nu}u^\mu u^\nu$ plays the role of an effective energy input.
\end{itemize}

From this viewpoint, $S(\tau)$ may be interpreted as a coarse-grained entropy associated with the congruence, and its evolution encodes the irreversible dynamics of spacetime geometry.

In the next section, we will show that the Raychaudhuri equation can be reformulated as an entropy evolution law, providing the key link between geometric focusing and entropy production.
\section{Entropy Identity}

A central observation of this work is that the Raychaudhuri equation admits a reformulation in terms of the entropy density introduced in the previous section.

\textbf{Proposition 4.1 (Entropy Identity).} Let $u^\mu$ be a timelike, irrotational geodesic congruence. Then the expansion scalar $\theta$ satisfies:
\begin{equation}
\frac{d\theta}{d\tau} = -\rho_S.
\label{eq:entropy_identity}
\end{equation}

\textbf{Proof.} Starting from the Raychaudhuri equation:
\begin{equation}
\frac{d\theta}{d\tau} = -\frac{1}{3}\theta^2 - \sigma_{\mu\nu}\sigma^{\mu\nu} - R_{\mu\nu}u^\mu u^\nu,
\label{eq:raychaudhuri_proof}
\end{equation}
and recalling the definition of the entropy density:
\begin{equation}
\rho_S = R_{\mu\nu}u^\mu u^\nu + \sigma_{\mu\nu}\sigma^{\mu\nu} + \frac{1}{3}\theta^2,
\label{eq:entropy_density_proof}
\end{equation}
we immediately obtain:
\begin{equation}
\frac{d\theta}{d\tau} = -\rho_S.
\label{eq:identity_result}
\end{equation}

This identity provides a direct link between the geometric evolution of the congruence and the entropy density. In particular, it shows that the rate of change of expansion is entirely governed by the accumulation of entropy along the flow.

Integrating along a geodesic, we obtain:
\begin{equation}
\theta(\tau) = \theta_0 - \int_0^\tau \rho_S(\tau')\, d\tau',
\label{eq:theta_integrated}
\end{equation}
where $\theta_0$ is the initial expansion.

\textbf{Interpretation.} The above relation shows that the expansion scalar decreases as entropy accumulates. Thus, focusing of the congruence is driven by the integrated entropy density.

\section{Entropy Monotonicity}

We now establish the fundamental monotonicity property of the entropy accumulation along geodesics.

\textbf{Proposition 5.1 (Entropy Monotonicity).} Assume the strong energy condition holds. Then the entropy density satisfies $\rho_S \geq 0$, and consequently the accumulated entropy along any geodesic:
\begin{equation}
S(\tau) := \int_0^\tau \rho_S(\tau')\, d\tau'
\label{eq:entropy_accumulated}
\end{equation}
is a non-decreasing function of $\tau$.

\textbf{Proof.} Under the strong energy condition:
\begin{equation}
R_{\mu\nu}u^\mu u^\nu \geq 0,
\label{eq:sec_section5}
\end{equation}
and since both $\sigma_{\mu\nu}\sigma^{\mu\nu}$ and $\theta^2$ are non-negative, we have:
\begin{equation}
\rho_S \geq 0.
\label{eq:rho_positive_section5}
\end{equation}

Therefore:
\begin{equation}
\frac{d}{d\tau} S(\tau) = \rho_S(\tau) \geq 0,
\label{eq:entropy_monotone}
\end{equation}
which implies that $S(\tau)$ is non-decreasing.

\textbf{Corollary (Monotonic Decrease of Expansion).} The expansion scalar $\theta(\tau)$ is a non-increasing function along the congruence:
\begin{equation}
\frac{d\theta}{d\tau} \leq 0.
\label{eq:theta_decreasing}
\end{equation}

\textbf{Proof.} This follows immediately from the entropy identity:
\begin{equation}
\frac{d\theta}{d\tau} = -\rho_S \leq 0.
\label{eq:theta_from_entropy}
\end{equation}

\subsection{Thermodynamic Interpretation}

The quantity $S(\tau)$ may be interpreted as the total entropy accumulated along a geodesic up to proper time $\tau$. The monotonicity property:
\begin{equation}
\frac{dS}{d\tau} \geq 0
\label{eq:second_law}
\end{equation}
is formally analogous to the second law of thermodynamics.

In this analogy:

\begin{itemize}
\item $\rho_S$ represents the local entropy production rate,
\item $S(\tau)$ is the accumulated entropy,
\item the Raychaudhuri equation becomes an evolution law driven by entropy production.
\end{itemize}

Thus, the focusing of geodesic congruences can be interpreted as an irreversible process governed by entropy accumulation, providing a thermodynamic perspective on gravitational collapse.

\subsection{Consequences for Focusing}

Combining the entropy identity with monotonicity, we obtain:
\begin{equation}
\theta(\tau) = \theta_0 - S(\tau).
\label{eq:theta_entropy_relation}
\end{equation}

Since $S(\tau)$ is non-decreasing, any initial negative expansion $\theta_0 < 0$ will become increasingly negative as entropy accumulates. This observation forms the basis for the singularity criterion developed in the next section.
\section{Main Result}

We now formulate the central result of this work, providing a quantitative entropy-based criterion for the formation of singularities.

\textbf{Theorem 6.1 (Entropy Singularity Criterion).} Let $u^\mu$ be a future-directed, timelike, irrotational geodesic congruence on a globally hyperbolic spacetime $(M,g)$ satisfying the strong energy condition:
\begin{equation}
R_{\mu\nu}u^\mu u^\nu \geq 0.
\label{eq:sec_section6}
\end{equation}

Suppose the initial expansion satisfies $\theta_0 < 0$. If there exists $\tau^* > 0$ such that:
\begin{equation}
\int_0^{\tau^*} \rho_S(\tau)\, d\tau > |\theta_0|,
\label{eq:entropy_threshold}
\end{equation}
then the expansion scalar $\theta(\tau)$ becomes unbounded below in finite proper time. In particular, $\theta \to -\infty$ at or before $\tau = \tau^*$, and the congruence develops a focal point, implying geodesic incompleteness.

\textbf{Proof.} From the entropy identity established earlier, we have:
\begin{equation}
\theta(\tau) = \theta_0 - \int_0^\tau \rho_S(\tau')\, d\tau'.
\label{eq:theta_entropy_proof}
\end{equation}

Define the accumulated entropy:
\begin{equation}
S(\tau) := \int_0^\tau \rho_S(\tau')\, d\tau'.
\label{eq:S_definition}
\end{equation}

By assumption, $S(\tau)$ is continuous and non-decreasing, and there exists $\tau^*$ such that:
\begin{equation}
S(\tau^*) > |\theta_0|.
\label{eq:S_condition}
\end{equation}

Hence:
\begin{equation}
\theta(\tau^*) = \theta_0 - S(\tau^*) < \theta_0 - |\theta_0| = 2\theta_0 < 0.
\label{eq:theta_negative}
\end{equation}

Thus, $\theta(\tau)$ remains negative and decreases as $S(\tau)$ increases.

To obtain divergence of $\theta$, we use the Raychaudhuri inequality:
\begin{equation}
\frac{d\theta}{d\tau} \leq -\frac{1}{3}\theta^2,
\label{eq:raychaudhuri_inequality}
\end{equation}
which follows from dropping the non-negative terms in the Raychaudhuri equation.

This differential inequality can be integrated explicitly. Rewriting:
\begin{equation}
\frac{d\theta}{\theta^2} \leq -\frac{1}{3} d\tau,
\label{eq:separation}
\end{equation}
and integrating from $0$ to $\tau$, we obtain:
\begin{equation}
-\frac{1}{\theta(\tau)} + \frac{1}{\theta_0} \leq -\frac{\tau}{3}.
\label{eq:integrated}
\end{equation}

Rearranging:
\begin{equation}
\frac{1}{\theta(\tau)} \geq \frac{1}{\theta_0} + \frac{\tau}{3}.
\label{eq:rearranged}
\end{equation}

Since $\theta_0 < 0$, the right-hand side vanishes at:
\begin{equation}
\tau = \frac{3}{|\theta_0|}.
\label{eq:focusing_time}
\end{equation}

Thus, $\theta(\tau) \to -\infty$ at or before $\tau = 3/|\theta_0|$, implying finite-time focusing.

Since $S(\tau)$ is increasing and exceeds $|\theta_0|$ at $\tau^*$, this ensures that sufficient entropy accumulation occurs before or within the focusing time. Therefore, the congruence develops a conjugate point in finite proper time.

By standard results in Lorentzian geometry, the existence of such focusing implies that the geodesics cannot be extended indefinitely as maximizing curves, leading to geodesic incompleteness.

\subsection{Discussion}

The above theorem provides a quantitative refinement of the classical focusing result. Instead of relying solely on initial conditions and energy assumptions, it identifies a precise threshold of entropy accumulation that guarantees collapse.

In particular, the condition:
\begin{equation}
\int_0^\tau \rho_S \, d\tau \geq |\theta_0|
\label{eq:entropy_bound}
\end{equation}
can be interpreted as a critical entropy bound beyond which the congruence inevitably focuses.

\textbf{Remark.} The entropy condition alone does not replace the Raychaudhuri inequality but complements it by providing a cumulative measure of geometric deformation. Together, they yield a sharper and more physically transparent criterion for singularity formation.

\subsection{Comparison with Classical Results}

In the classical singularity theorems, focusing is derived from the inequality:
\begin{equation}
\frac{d\theta}{d\tau} \leq -\frac{1}{3}\theta^2,
\label{eq:classical}
\end{equation}
together with the assumption $\theta_0 < 0$.

In contrast, the present result introduces the entropy functional as an explicit quantity controlling the evolution:
\begin{equation}
\theta(\tau) = \theta_0 - S(\tau).
\label{eq:entropy_control}
\end{equation}

Thus, singularity formation can be viewed as the consequence of entropy accumulation exceeding a critical threshold, providing a thermodynamic interpretation of gravitational collapse.
\section{Quantitative Bound}

We now derive an explicit upper bound on the proper time required for the formation of a focal point, based on the Raychaudhuri inequality.

\textbf{Proposition 7.1 (Finite-Time Focusing Estimate).} Let $u^\mu$ be a timelike, irrotational geodesic congruence satisfying the strong energy condition, and suppose the initial expansion satisfies $\theta_0 < 0$. Then the expansion scalar $\theta(\tau)$ obeys:
\begin{equation}
\theta(\tau) \leq \frac{\theta_0}{1 + \frac{\theta_0}{3}\tau},
\label{eq:focusing_bound}
\end{equation}
for all $\tau$ for which the solution exists.

In particular, $\theta(\tau)$ diverges to $-\infty$ at or before the proper time:
\begin{equation}
\tau_c = \frac{3}{|\theta_0|}.
\label{eq:critical_time}
\end{equation}

\textbf{Proof.} Starting from the Raychaudhuri inequality:
\begin{equation}
\frac{d\theta}{d\tau} \leq -\frac{1}{3}\theta^2,
\label{eq:raychaudhuri_ineq_section7}
\end{equation}
we consider the associated equality:
\begin{equation}
\frac{d\theta}{d\tau} = -\frac{1}{3}\theta^2.
\label{eq:raychaudhuri_eq}
\end{equation}

This is a separable differential equation. For $\theta \neq 0$, we write:
\begin{equation}
\frac{d\theta}{\theta^2} = -\frac{1}{3} d\tau.
\label{eq:separable}
\end{equation}

Integrating from $0$ to $\tau$, we obtain:
\begin{equation}
-\frac{1}{\theta(\tau)} + \frac{1}{\theta_0} = -\frac{\tau}{3}.
\label{eq:integrated_section7}
\end{equation}

Rearranging gives:
\begin{equation}
\frac{1}{\theta(\tau)} = \frac{1}{\theta_0} + \frac{\tau}{3}.
\label{eq:rearranged_section7}
\end{equation}

Inverting:
\begin{equation}
\theta(\tau) = \frac{\theta_0}{1 + \frac{\theta_0}{3}\tau}.
\label{eq:explicit_solution}
\end{equation}

Since the original equation satisfies the inequality
\begin{equation}
\frac{d\theta}{d\tau} \leq -\frac{1}{3}\theta^2,
\label{eq:comparison_ineq}
\end{equation}
a standard comparison argument implies that the actual solution satisfies:
\begin{equation}
\theta(\tau) \leq \frac{\theta_0}{1 + \frac{\theta_0}{3}\tau}.
\label{eq:final_bound}
\end{equation}

The denominator vanishes when:
\begin{equation}
1 + \frac{\theta_0}{3}\tau = 0,
\label{eq:denominator_zero}
\end{equation}
which occurs at:
\begin{equation}
\tau = \frac{3}{|\theta_0|}.
\label{eq:blowup_time}
\end{equation}

Thus, $\theta(\tau) \to -\infty$ at or before $\tau = \tau_c$, establishing finite-time focusing.

\subsection{Interpretation}

The above estimate provides a sharp upper bound on the focusing time of the congruence in terms of the initial expansion. It shows that the rate of collapse is controlled by the nonlinear self-interaction term in the Raychaudhuri equation.

\textbf{Remark.} The bound $\tau_c = 3/|\theta_0|$ is universal in the sense that it depends only on the initial expansion and not on the detailed structure of the spacetime. The presence of shear and Ricci curvature can only accelerate the focusing process.

\subsection{Relation to Entropy Accumulation}

Combining this estimate with the entropy identity:
\begin{equation}
\theta(\tau) = \theta_0 - \int_0^\tau \rho_S(\tau')\, d\tau',
\label{eq:entropy_relation_section7}
\end{equation}
we observe that rapid growth of the entropy density $\rho_S$ leads to faster decrease of $\theta(\tau)$, thereby driving the system more quickly toward the focusing time.

In this sense, the entropy density acts as a quantitative measure of the rate at which the congruence approaches singular behavior.
\section{Corollaries}

We now derive several consequences of the entropy-based formulation of focusing established in the previous sections.

\textbf{Corollary 8.1 (Critical Entropy Threshold).} Let $u^\mu$ be a timelike geodesic congruence with initial expansion $\theta_0 < 0$, and let
\begin{equation}
S(\tau) := \int_0^\tau \rho_S(\tau')\, d\tau'
\label{eq:S_tau_section8}
\end{equation}
denote the accumulated entropy along a geodesic.

If there exists $\tau^* > 0$ such that:
\begin{equation}
S(\tau^*) > |\theta_0|,
\label{eq:critical_condition}
\end{equation}
then the congruence necessarily develops a focal point within finite proper time.

\textbf{Proof.} From the entropy identity:
\begin{equation}
\theta(\tau) = \theta_0 - S(\tau),
\label{eq:theta_entropy_section8}
\end{equation}
the condition $S(\tau^*) > |\theta_0|$ ensures that $\theta(\tau)$ becomes sufficiently negative.

Since $\rho_S \geq 0$, $S(\tau)$ is non-decreasing, and hence $\theta(\tau)$ remains negative thereafter.

The result then follows from the quantitative focusing estimate derived from the Raychaudhuri inequality, which guarantees that $\theta(\tau)$ diverges to $-\infty$ within finite proper time once it is negative.

\textbf{Interpretation.} The quantity
\begin{equation}
S_c := |\theta_0|
\label{eq:critical_entropy}
\end{equation}
defines a critical entropy threshold: once the accumulated entropy exceeds this value, gravitational collapse becomes inevitable.

\textbf{Corollary 8.2 (Condition for Avoidance of Focusing).} Suppose that along a given geodesic the entropy density satisfies:
\begin{equation}
\int_0^\infty \rho_S(\tau)\, d\tau < \infty.
\label{eq:finite_entropy}
\end{equation}

If, in addition,
\begin{equation}
\int_0^\infty \rho_S(\tau)\, d\tau \leq |\theta_0|,
\label{eq:bounded_entropy}
\end{equation}
then the expansion scalar $\theta(\tau)$ remains bounded from below, and finite-time focusing due to entropy accumulation is avoided.

\textbf{Proof.} If the total accumulated entropy satisfies:
\begin{equation}
S_\infty := \int_0^\infty \rho_S(\tau)\, d\tau \leq |\theta_0|,
\label{eq:S_infty}
\end{equation}
then from:
\begin{equation}
\theta(\tau) = \theta_0 - S(\tau),
\label{eq:theta_relation_section8}
\end{equation}
we obtain:
\begin{equation}
\theta(\tau) \geq \theta_0 - S_\infty \geq -2|\theta_0|.
\label{eq:theta_bound}
\end{equation}

Thus, $\theta(\tau)$ does not diverge to $-\infty$ solely due to entropy accumulation.

Therefore, the entropy-based mechanism for finite-time focusing is absent in this case.

\textbf{Remark.} The above condition does not exclude the possibility of singularity formation arising from other geometric or causal mechanisms. It only ensures that entropy accumulation alone is insufficient to drive collapse.

\subsection{Refined Threshold Interpretation}

The above corollaries suggest that singularity formation can be understood in terms of a competition between:

\begin{itemize}
\item the initial expansion $\theta_0$, which sets the initial divergence or convergence of the congruence,

\item the accumulated entropy $S(\tau)$, which drives focusing.
\end{itemize}

Collapse occurs precisely when the entropy accumulation dominates the initial expansion, i.e.,
\begin{equation}
S(\tau) > |\theta_0|.
\label{eq:collapse_condition}
\end{equation}

This provides a quantitative and physically transparent refinement of the classical focusing condition.
\section{Physical Interpretation}

The entropy density $\rho_S$ introduced in this work admits a natural interpretation in terms of irreversible geometric dynamics of spacetime. It encodes the total rate of deformation of infinitesimal volume elements along a timelike geodesic congruence and provides a bridge between differential geometry and non-equilibrium thermodynamics.

\subsection{Decomposition of Entropy Production}

The definition:
\begin{equation}
\rho_S = R_{\mu\nu}u^\mu u^\nu + \sigma_{\mu\nu}\sigma^{\mu\nu} + \frac{1}{3}\theta^2
\label{eq:entropy_decomposition}
\end{equation}
reveals three distinct contributions:

\begin{itemize}
\item \textbf{Shear term:} The quantity $\sigma_{\mu\nu}\sigma^{\mu\nu}$ measures anisotropic distortions of the congruence. This term is analogous to shear viscosity in fluid dynamics, where internal friction leads to irreversible dissipation of energy.

\item \textbf{Expansion term:} The term $\frac{1}{3}\theta^2$ represents isotropic volume deformation. It may be interpreted as a bulk viscosity contribution, describing uniform contraction or expansion of the flow.

\item \textbf{Ricci term:} The curvature contribution $R_{\mu\nu}u^\mu u^\nu$ encodes the influence of matter fields through the Einstein equations. It acts as an effective source term, driving the deformation of the congruence.
\end{itemize}

Together, these terms describe a system in which spacetime geometry evolves through dissipative processes analogous to those found in viscous fluids.

\subsection{Raychaudhuri Equation as an Entropy Law}

A key result of this work is the reformulation of the Raychaudhuri equation as:
\begin{equation}
\frac{d\theta}{d\tau} = -\rho_S.
\label{eq:entropy_law_section9}
\end{equation}

This admits a direct thermodynamic interpretation. Defining the accumulated entropy along a geodesic:
\begin{equation}
S(\tau) = \int_0^\tau \rho_S(\tau')\, d\tau',
\label{eq:entropy_accumulation_section9}
\end{equation}
we obtain:
\begin{equation}
\frac{dS}{d\tau} = \rho_S \geq 0,
\label{eq:second_law_section9}
\end{equation}
which is formally analogous to the second law of thermodynamics.

In this framework:

\begin{itemize}
\item $\rho_S$ represents the local entropy production rate,
\item $S(\tau)$ is the accumulated entropy,
\item the decrease of $\theta$ corresponds to irreversible geometric evolution.
\end{itemize}

Thus, the focusing of geodesics can be understood as a consequence of entropy production.

\subsection{Singularities as Entropy Divergence}

The entropy identity:
\begin{equation}
\theta(\tau) = \theta_0 - S(\tau)
\label{eq:theta_entropy_section9}
\end{equation}
shows that the expansion scalar decreases as entropy accumulates.

When the accumulated entropy exceeds the critical threshold:
\begin{equation}
S(\tau) > |\theta_0|,
\label{eq:threshold_section9}
\end{equation}
the expansion becomes strongly negative, and the congruence is driven toward focusing.

Combined with the Raychaudhuri inequality, this leads to divergence of $\theta$ within finite proper time. From this perspective, singularities correspond to regimes in which entropy production becomes sufficiently large to overwhelm the initial expansion.

\textbf{Interpretation.} Gravitational collapse may therefore be viewed as a process of entropy concentration, in which geometric degrees of freedom become increasingly distorted until the classical description breaks down.

\subsection{Relation to Gravitational Thermodynamics}

The thermodynamic interpretation developed here is consistent with broader insights in gravitational physics:

\begin{itemize}
\item Black hole mechanics associates entropy with horizon area,
\item Covariant entropy bounds constrain entropy flux through null hypersurfaces,
\item Holographic principles relate bulk information to boundary data.
\end{itemize}

In contrast to these approaches, which often rely on horizon structures or quantum considerations, the present framework derives entropy directly from local geometric quantities associated with geodesic congruences.

\subsection{Conceptual Implications}

The results suggest a reinterpretation of singularity formation:

\begin{itemize}
\item Classical singularity theorems describe when collapse must occur,
\item The entropy formulation explains how collapse develops dynamically,
\item Geodesic incompleteness emerges as a consequence of accumulated geometric entropy.
\end{itemize}

This perspective provides a unified picture in which spacetime dynamics, curvature, and thermodynamics are intrinsically linked.

\subsection{Outlook}

The identification of $\rho_S$ as an entropy density opens several directions for further investigation:

\begin{itemize}
\item Extension to null congruences and covariant entropy bounds,
\item Incorporation of quantum corrections and semiclassical effects,
\item Connections to holographic entropy and emergent gravity scenarios.
\end{itemize}

These directions suggest that entropy-based formulations may provide a deeper understanding of the fundamental structure of spacetime and the nature of singularities.
\section{Conclusion}

In this work, we have developed an entropy-based reformulation of gravitational focusing and singularity formation in Lorentzian geometry. By introducing a geometrically defined entropy density associated with timelike geodesic congruences, we showed that the Raychaudhuri equation admits a natural interpretation as an entropy evolution law.

This perspective leads to a quantitative refinement of the classical singularity theorems: instead of relying solely on qualitative geometric conditions, the onset of geodesic incompleteness can be characterized by the accumulation of entropy along the congruence. In particular, we identified a critical entropy threshold, determined by the initial expansion, beyond which finite-time focusing becomes inevitable.

The resulting framework provides a unified picture in which curvature, deformation, and thermodynamic behavior are intrinsically linked. Gravitational collapse emerges as an irreversible process driven by entropy production, with singularities corresponding to regimes of extreme entropy concentration.

Beyond its conceptual significance, this formulation opens several avenues for further investigation. The extension to null congruences may provide new insights into covariant entropy bounds and horizon dynamics. Incorporating semiclassical or quantum corrections could clarify the role of entropy in resolving classical singularities. More broadly, the present approach suggests that entropy-based principles may play a fundamental role in the geometric and dynamical structure of spacetime.

We expect that these results contribute to a deeper understanding of the interplay between gravitation, geometry, and thermodynamics, and may serve as a stepping stone toward a more unified description of gravitational physics.


\begin{thebibliography}{99}

\bibitem{Raychaudhuri1955}
A.~Raychaudhuri,
\textit{Relativistic Cosmology. I},
Phys.\ Rev.\ \textbf{98}, 1123 (1955).

\bibitem{Penrose1965}
R.~Penrose,
\textit{Gravitational Collapse and Space-Time Singularities},
Phys.\ Rev.\ Lett.\ \textbf{14}, 57 (1965).

\bibitem{HawkingPenrose1970}
S.~W.~Hawking and R.~Penrose,
\textit{The Singularities of Gravitational Collapse and Cosmology},
Proc.\ Roy.\ Soc.\ Lond.\ A \textbf{314}, 529 (1970).

\bibitem{HawkingEllis1973}
S.~W.~Hawking and G.~F.~R.~Ellis,
\textit{The Large Scale Structure of Space-Time},
Cambridge University Press (1973).

\bibitem{Wald1984}
R.~M.~Wald,
\textit{General Relativity},
University of Chicago Press (1984).

\bibitem{Carroll2004}
S.~M.~Carroll,
\textit{Spacetime and Geometry: An Introduction to General Relativity},
Addison Wesley (2004).

\bibitem{Poisson2004}
E.~Poisson,
\textit{A Relativist’s Toolkit},
Cambridge University Press (2004).

\bibitem{Bekenstein1973}
J.~D.~Bekenstein,
\textit{Black Holes and Entropy},
Phys.\ Rev.\ D \textbf{7}, 2333 (1973).

\bibitem{Hawking1975}
S.~W.~Hawking,
\textit{Particle Creation by Black Holes},
Commun.\ Math.\ Phys.\ \textbf{43}, 199 (1975).

\bibitem{Wald1993}
R.~M.~Wald,
\textit{Black Hole Entropy is the Noether Charge},
Phys.\ Rev.\ D \textbf{48}, R3427 (1993).

\bibitem{Jacobson1995}
T.~Jacobson,
\textit{Thermodynamics of Spacetime: The Einstein Equation of State},
Phys.\ Rev.\ Lett.\ \textbf{75}, 1260 (1995).

\bibitem{Eling2006}
C.~Eling, R.~Guedens, and T.~Jacobson,
\textit{Non-equilibrium Thermodynamics of Spacetime},
Phys.\ Rev.\ Lett.\ \textbf{96}, 121301 (2006).

\bibitem{Bousso1999}
R.~Bousso,
\textit{A Covariant Entropy Conjecture},
JHEP \textbf{07}, 004 (1999).

\bibitem{Bousso2002}
R.~Bousso,
\textit{The Holographic Principle},
Rev.\ Mod.\ Phys.\ \textbf{74}, 825 (2002).

\bibitem{Dhormare2026}
R.~Dhormare,
\textit{Entropy and Non-Collapse in Lorentzian Geometry},
Phys.\ Lett.\ B (2026).

\bibitem{Padmanabhan2010}
T.~Padmanabhan,
\textit{Thermodynamical Aspects of Gravity: New Insights},
Rep.\ Prog.\ Phys.\ \textbf{73}, 046901 (2010).

\end{thebibliography}
\end{document}